\documentclass[%
 reprint,
 amsmath,amssymb,
 aps,
 nofootinbib,
 superscriptaddress
]{revtex4-1}

\usepackage{graphicx}% Include figure files
\usepackage{dcolumn}% Align table columns on decimal point
\usepackage{bm}% bold math
\usepackage{enumitem}
\usepackage[draft, inline, nomargin]{fixme}
\fxsetup{theme=color, mode=multiuser}
\FXRegisterAuthor{ab}{1}{\color{blue}AB}
\FXRegisterAuthor{fs}{2}{\color{red}FS}
\FXRegisterAuthor{jx}{3}{\color{green}JX}
\FXRegisterAuthor{sp}{4}{\color{cyan}SP}
\usepackage{lineno}
\usepackage{hyperref}

\newcommand{\us}{$\mu$s}
\newcommand{\dl}{DAMA/LIBRA }
\newcommand{\nai}{NaI(Tl)}

\begin{document}
%\linenumbers

%\preprint{APS/123-QED}

\title{Production and suppression of delayed light in NaI}
%UV-induced delayed light emission in NaI and a method to suppress it} %Energy storage and release in sodium iodide scintillation crystals}

\author{F.~Sutanto} \email{sutanto2@llnl.gov}
\author{J.~Xu}
\author{S.~Pereverzev}
\author{A.~Bernstein}
\affiliation{Lawrence Livermore National Laboratory, Livermore, CA 94550, USA}

\date{\today}

\begin{abstract}
\noindent We investigate a hypothesis that energy accumulation and the subsequent release in \nai\ may lead to pulse-like events in the few-keV energy regime, a phenomenon suggested by the crystal manufacturing company Saint-Gobain, who provided the crystals for \dl. While we observed delayed long-lasting light emission in a 3” \nai\ crystal after exposing it to UV light, the delayed light consists primarily of single photons that are uncorrelated with each other. We also observe delayed light emission in \nai\ following gamma radiation and large ionization events like cosmic-ray muons. We found that irradiating the crystal with red light after UV exposure significantly suppressed delayed photon emissions. 
\end{abstract}

\pacs{Valid PACS appear here}
\maketitle

\section{Introduction}

In the last few decades, the quest for dark matter (DM) has given a null result but for a few exceptions~\cite{cogent_2011,cogent_2014,cresst_2012,dama-libra-phase2-first-six-cycle}.  
%several direct-detection experiments have reported possible hints of dark matter (DM) interactions~\cite{cogent_2011,cogent_2014,cresst_2012,dama-libra-phase2-first-six-cycle}.  
To this date, the \dl\ collaboration maintains the sole remaining claim~\cite{dama-libra-phase2-first-six-cycle,dama,dama-libra-nai,dama-libra-phase1,dama-libra-phase2},  which has been in strong tension with other experiments~\cite{dama-libra_exclusion,cosine100_nature,cosine100_prl,cosine100_sciAdv,anais_status_2016,anais-112_threeYearsResult,anais-112_twoYearsResult,dm-ice_2015}.
The \dl\ detector is situated in the Laboratori Nazionali del Gran Sasso (LNGS) in Italy at an overburden of 1400~m~\cite{dama-libra-apparatus}.
The active volume consists of an array of ultra-high purity NaI(Tl) crystals with a total active mass of 250~kg~\cite{dama-libra-nai,dama-libra-apparatus}, and passively shielded with high-purity copper, lead, polyethylene, and concrete layers. 
The dark matter claim was made based on an event rate modulation observed in the low energy region of the experimental spectrum. %in the low-energy region of the experiment.
A model-independent analysis of twenty years of \dl\ data reports a modulation of 2--6~keVee event rate over 14 annual cycles with 9.3 sigma detection level~\cite{dama-libra-phase1}.
Reduction of the energy threshold to 1~keVee in the latest six annual cycles improves the confirmation of the annual modulation at a 12.9 sigma detection level~\cite{dama-libra-phase2-first-six-cycle}. 
Recently, combined data of DAMA/NaI, DAMA/LIBRA-phase1, and DAMA/LIBRA-phase2 favor the presence of a modulated behavior of 2--6~keVee event rate at a 13.7 sigma confidence level~\cite{dama-libra-13p7sigma-2-6keV}.
A few other NaI-based experiments have been launched with the goal of performing a model-independent test of the \dl claim~\cite{cosine100_nature,cosine100_prl,cosine100_sciAdv,anais_status_2016,anais-112_threeYearsResult,anais-112_twoYearsResult,dm-ice_2015}.
Among all presently running NaI-based experiments except DAMA, the SABRE detector demonstrates the lowest background rate (1.20$\pm$0.05 cpd/kg/keV) in the 1--6~keV energy range~\cite{sabre_2021}. However, no experiments have obtained a background rate as low as that in \dl\ yet. 
As a result, the \dl\ annual modulation has not been definitively confirmed or refuted up to date. 

Other authors have sought to explain the \dl\ results via alternative hypotheses based on known physics processes.
For example, it has been demonstrated that a slowly varying time-dependent background can result in an annual modulation when data are treated in terms of residuals obtained by subtracting the yearly average background such as that used in the \dl\ analysis~\cite{buttazzo_2020}. 
Analysis of COSINE-100 data ignoring information from the veto detectors and using \dl\ analysis methods shows a slowly decreasing background partly due to the decay of cosmogenically activated nuclides. 
However, the residual rate obtained by subtracting the yearly average background shows an opposite modulation phase to \dl\ \cite{cosine_2022}.
It was also proposed that the seasonal modulation of cosmogenic background could explain the observed \dl\ signal~\cite{neutron_to_explain_damalibra_0,neutron_to_explain_damalibra_1,neutron_to_explain_damalibra_2}.
However, the direct interaction rates of such processes in the few-keV energy region are estimated to fall short by orders of magnitude compared to the observed \dl\ low-energy event rate~\cite{muonToExplainDL_0,muonToExplainDL_1,dm-ice_muon}.
The presence of radioactive impurities has also been proposed to explain the observed low-energy event rate and may have the potential to modulate over time from event selection criteria.
However, \dl\ claimed very low levels of radioactive contaminants~\cite{DL_impurityLevel} that are not sufficient to offer a viable explanation for the observed low-energy event rate~\cite{impurityToExplainDL_rad}.
The claimed impurity contamination level also rules out the potential of DM scattering on impurities as an explanation for the observed modulation~\cite{impurityToExplainDL_OH,impurityToExplainDL_Ar,impurityToExplainDL_Tl}.

%A relatively recent 
One hypothesis suggests that non-radioactive processes such as delayed photon emission as a result of accumulated energy discharge may account for the claimed DM signal in the \dl\ experiment~\cite{nygren_2011}.
While no specific mechanism leading to such relaxation was offered, the authors noted that long-lasting photon production in \nai\ has been observed previously~\cite{nygren_2011}.
Saint-Gobain, the supplier of \nai\ crystals for the \dl\ experiment, reports that~\cite{stGobain}: 
\begin{quote}
    With mild UV exposure, several pulses per second can be seen in the 6-10~keV region of a spectrum. If the crystal is stored in a dark area, this mild UV exposure will eventually disappear, although it may take from several hours to several days for the effects to stop.
\end{quote}

In order for energy discharge following accumulation and storage to explain the \dl\ observation, the photons emitted need to match that of NaI scintillation to effectively pass the \dl\ event selection criteria.
A mechanism known as Self Organized Criticality (SOC)~\cite{soc_bak2} offers a feasible model for how such pulses might develop following a gradual accumulation of energy~\cite{sergey_2021}. In this hypothesis, small amounts of energy gradually accumulate and are stored in excitonic or other states within the crystal, which can collectively relax and release the energy in an avalanche-like process on a time scale much shorter than that required for energy accumulation. 
If this hypothesis can be confirmed in \nai, it may also provide hints on the recurring background patterns seen in many dark matter detectors, i.e., the sharp rise of event rate at the energy region approaching those of excitations in the detector materials~\cite{sharpRise_cogent,sharpRise_darkSide,sharpRise_sensei,sharpRise_xenon,sharpRise_xenon100,sharpRise_xenon1t,sergey_2021,sharpRise_excess}.

Consistency with the \dl\ data also requires the release of the stored energy to exhibit seasonal modulation aligned with the observed annual peak in the \dl data. 
If muons were to be responsible for the storage and release of energy in \dl, an as-yet-unknown mechanism to shift the maximum production of such pulses forward from the seasonal maximum in muon flux by approximately 45~days would be needed~\cite{dm-ice_muon,hubbard_dissertation}. Environmental factors, such as changes in pressure, temperature, electric/magnetic fields,
mechanical stress, or vibration, might lead to the annual modulation of delayed light emission, but these effects will be out of the scope of the current study. 

Recognizing that a careful experimental study of delayed light emission processes in NaI is required, we sought to test the Saint-Gobain observations in a more controlled fashion than we could find in published literature. While additional assumptions would be needed for this delayed emission effect to explain the \dl\ low-energy signal modulation, the possibility of pulsed events with the characteristics of keV-scale recoils produced by energy accumulation would itself be a significant result. In this work, a 3" \nai\ crystal was exposed to UV light, MeV-scale gamma-rays, and muons. 
%or other forms of radiation such as $^{60}$Co irradiation and muon passages,
The detector response was then monitored for approximately two days following UV and gamma-ray exposure and a few hundred milliseconds following a muon trigger ~\cite{dm-ice_muon,hubbard_dissertation,delayEmission_co60,delayEmission_mu,delayEmission_radSource}, we observed delayed light emission following the different forms of irradiation of the \nai\ crystal.  However, our analysis suggests that the majority of the photons are uncorrelated with each other, and do not form correlated pulse-like events. 
Lacking the characteristic time structure of a typical \nai pulse (a correlated emission of light with a decay time constant of 200--300~ns), these uncorrelated events can largely be suppressed with pulse-shape-based cuts. However, leakage of these events into the \dl\ signal region remains a possibility. This experiment also does not rule out the presence of keV-scale pulses whose rate is too low to be seen in the presence of the  strong uncorrelated afterglow signal. 
%This experiment also does not rule out the possibility that keV-scale pulses are 
%produced by the energy storage with a low rate in \nai\ and are overwhelmed by the afterglow and therefore avoid being detected. 
We further demonstrated that exposure of the \nai\ crystal after UV radiation to red light reduces the intensity of the delayed photon emissions, providing an opportunity for \nai\ experiments and other low energy-oriented experiments to suppress potential parasitic backgrounds.

The remainder of this article is organized as follows. 
In Section~\ref{sec_materialAndMethod}, we describe the experimental setup and the data acquisition scheme.
The temporal characteristics of the long-lived photon signals are explained in Section~\ref{subsec_discussion_tenporal}. 
The possible underlying process leading to such relaxation is discussed, followed by a discussion on a technique to suppress the delayed light discovered in this work in Section~\ref{subsec_red}.
The relevance of our observation to the \dl\ experiment is discussed in Section~\ref{subsec_otherExp}.
We present our conclusion and propose future work in Section~\ref{sec_conclusion}.

%%%%%%%%%%%%%%%%%%%%%%%%%%%%%%%%%%%%%%%%%%%%%%%%%%%%%%%%%%%%%%%%%%%%%
\section{\label{sec_materialAndMethod}Experimental setup}
%%%%%%%%%%%%%%%%%%%%%%%%%%%%%%%%%%%%%%%%%%%%%%%%%%%%%%%%%%%%%%%%%%%%%

\begin{figure}
    \includegraphics[width=1.0\linewidth]{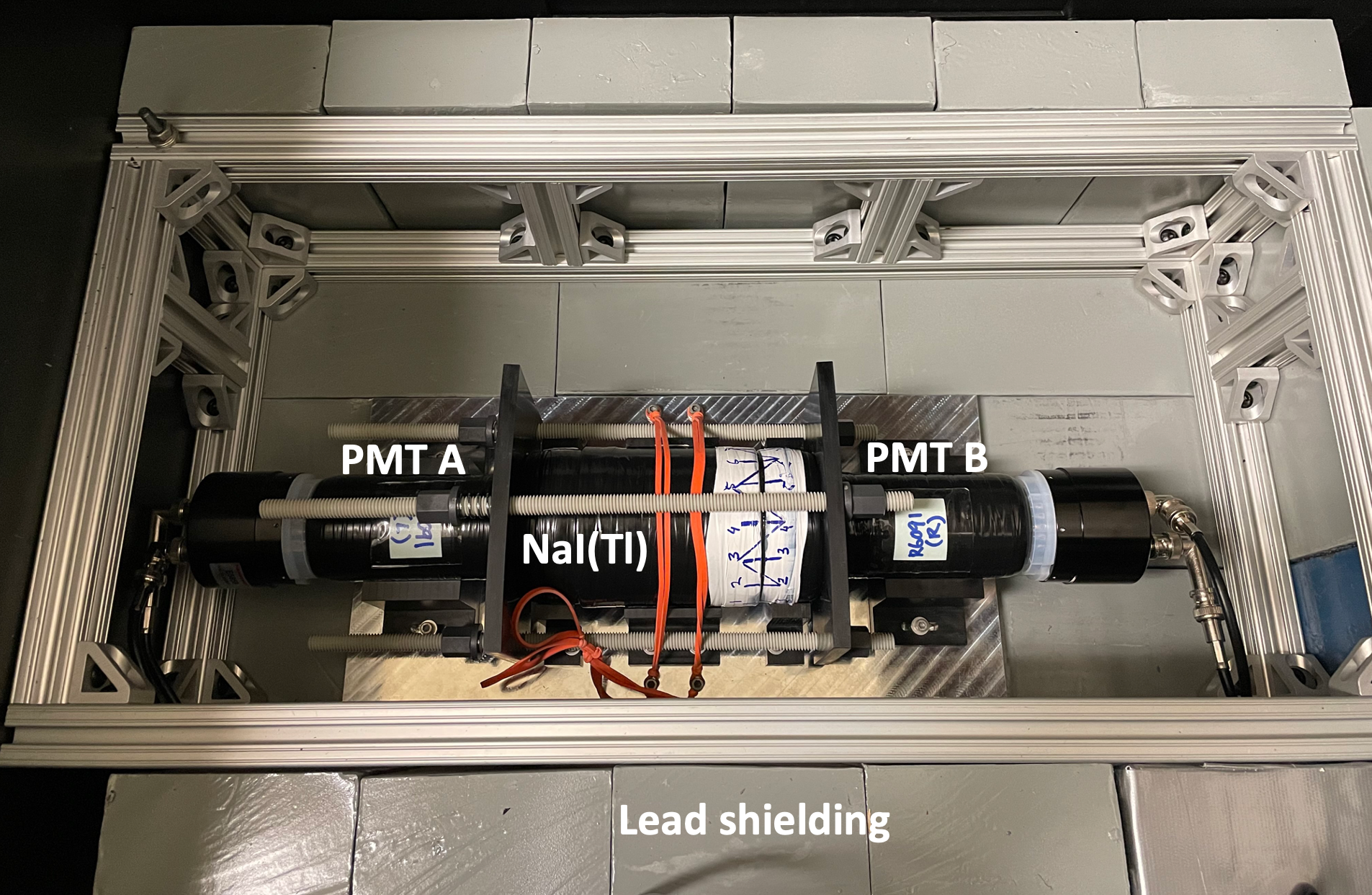}
    \caption{Experimental setup. The 3 in NaI(Tl) crystal was coupled to two Hamamatsu R6091 PMTs. The support structure was used to keep the detector and PMTs concentric. We used two-inch thick lead bricks to passively shield the target volume against %cosmic gamma-ray backgrounds and those originating from the surrounding materials.
    ambient gamma backgrounds.}
    \label{fig_naiSetup}
\end{figure}

\begin{figure}
    \includegraphics[width=1.0\linewidth]{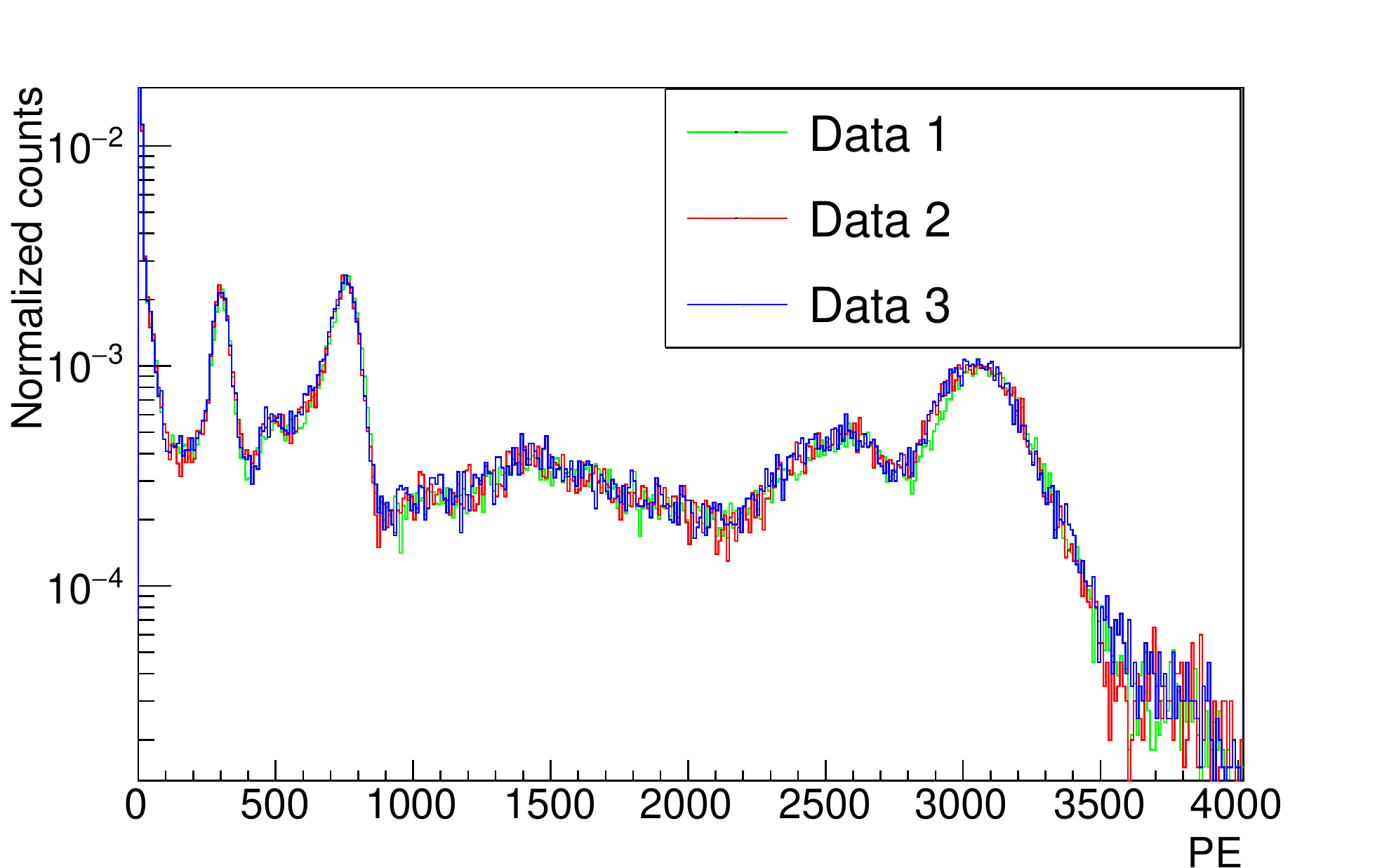}
    \caption{The detector response to $^{133}$Ba source with a normalized area under the curve. For each set of data, the support structure was disassembled and reassembled to test the reproducibility of our energy response.}
    \label{fig_ergCali}
\end{figure}

A 3'' $\times$ 3'' \nai\ crystal fabricated by Saint-Gobain~\cite{stGobain_nai_specSheet} was used in this experiment.
The crystal size was chosen to be sufficiently large compared to the few-centimeter UV absorption length~\cite{uv_absLength_nai} of NaI(Tl) crystals. Saint-Gobain also reported that the effect of UV exposure on \nai\ is most noticeable in large crystals~\cite{stGobain}. The crystal was encapsulated in a 0.02-inch thick cylindrical aluminum container with two 0.25-inch thick commercial-grade quartz glass optical windows to view the crystal. The quartz glass has $>$90\% transmission within [300,2000]~nm range, allowing passage of both scintillation and UV light with minimal losses. The inner cylindrical wall was coated with a proprietary white reflector to improve light collection. To ensure high light-collection efficiency, Eljen-550 optical silicone grease was used to couple the crystal module to two R6091 Hamamatsu head-on photomultiplier tubes that have high gain values (5$\times10^{6}$ at 1500~V) and good quantum efficiency ($\sim$26~\% at 420~nm) at the scintillation peak of NaI(Tl) crystal~\cite{r6091_specSheet}. The light yield of our detector was estimated to be 9.89~photoelectron/keVee (PE/keVee) from the 30.85~keV peak observed in the measured $^{133}$Ba spectrum (Fig.~\ref{fig_ergCali}).

\begin{figure}
    \includegraphics[width=1.0\linewidth]{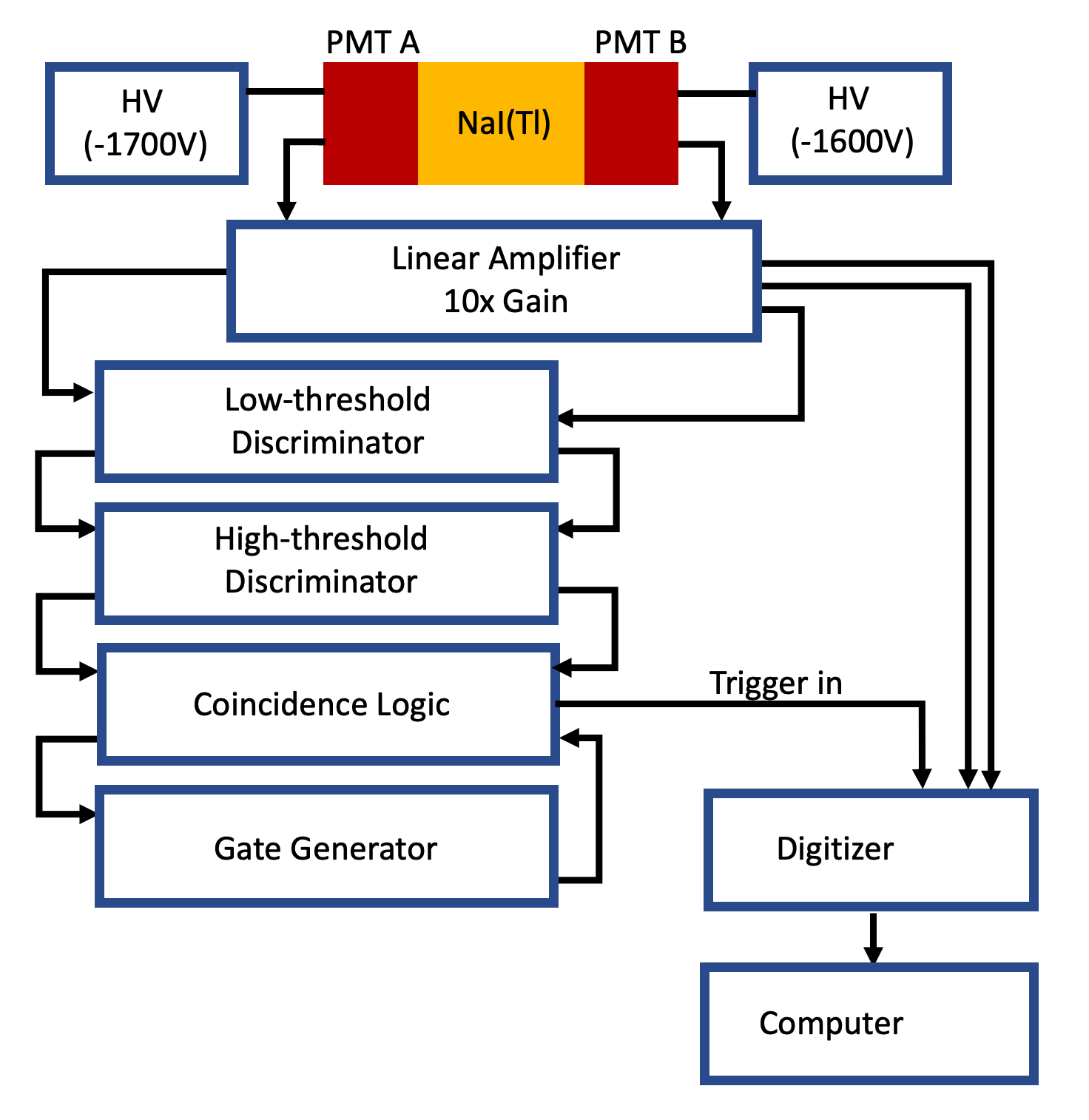}
    \caption{Coincidence trigger scheme used in this experiment. Each PMT signal was amplified by $\times$10 and split into two copies, with one copy digitized and the other used to generate a 1$\mu$s coincidence trigger between single photoelectron pulses in each PMT. }
    \label{fig_electronics}
\end{figure}

Figure~\ref{fig_naiSetup} shows the support structure used to keep the assembly concentric and maintain good optical contact between the crystal and the PMTs. The PMTs were wrapped in polytetrafluoroethylene (PTFE) tape (POLY-TEMP PN-16050, $>$98\% reflectivity) to maintain the consistency of the light collection. The crystal and the PMTs were supported by Delrin-based holders, which were secured to an aluminum base plate. Elastomer bands were used as an additional safety measure, holding the crystal in place. Two Delrin-based alignment plates, fiberglass rods, fiberglass flange nuts, and compression springs were used to apply moderate pressure between the detector and the PMT faceplates (Fig.~\ref{fig_naiSetup}). 
To suppress gamma-ray and other ambient backgrounds in low-energy regions, we placed two-inch-thick lead bricks on the side and below the detector and one-inch-thick lead bricks on the top of the detector to passively shield the active volume.
The support structure and the surrounding lead shielding were installed in a dark box.

Full-waveform digitization of the PMT outputs is chosen to provide an accurate determination of event energy, timing, and pulse shape. 
An ORTEC~556 high voltage power supply provides -1700 V and -1600 V to PMT A and PMT B, respectively. Each PMT signal was amplified using a CAEN~V975 unit (10$\times$-gain), producing two sets of outputs. One set was sent to a 16-channel 250-MHz 14-bit SIS3316 Struck digitizer to be recorded, and the other was sent to a series of discriminators and logic units to produce a trigger for the data acquisition system. 

In this experiment, we used two different trigger schemes: a coincidence trigger commonly used in NaI(Tl) detectors and a periodic trigger mode. 
The coincidence trigger scheme is illustrated in Fig.~\ref{fig_electronics}.
The amplified PMT outputs were first sent to an 8-channel
CAEN N840 low-threshold leading-edge discriminator, which uses a threshold of 5~mV to trigger on nearly all single photoelectron pulses while rejecting a large fraction of random noise. 
The low-threshold discriminator outputs were then sent to LeCroy~821 high-level discriminator (used as a fast gate generator) to produce 1~$\mu$s coincidence windows and an LRS~122 logic unit was used to generate a coincidence trigger.
For each coincidence trigger, waveforms were digitized 2~$\mu$s before and 5~$\mu$s after the trigger time.
To suppress repetitive triggers within the same event window, we also use one copy of the LRS~122 coincidence outputs to produce a 10-$\mu$s pulse in GG~8000 octal gate generator to veto the LRS~122 unit after each trigger. 
This arrangement ensures pulses recorded by the digitizer channels are well synchronized but also causes 3~$\mu$s of deadtime for each trigger. 
This trigger logic can detect kHz level of keV-like pulses with high efficiency, but has severe efficiency loss for recording uncorrelated photoelectron pulses following UV radiation. 

%A periodic trigger data acquisition mode was introduced to record PMT outputs near-continuously following UV exposure. 
In the periodic trigger mode, a forced periodic trigger was provided by a Tektronik AFG31000 pulse generator operating at 3600~Hz (278~$\mu$s period), with a 99.9\% duty cycle.
For each trigger, we recorded waveforms from 6~$\mu$s before the trigger time to 244~$\mu$s after. 
The 28~$\mu$s time gap between recorded waveforms is needed for the digitizer to store the previous waveform.
Both PMT waveforms are recorded separately, and a single p.e. in either is defined as a valid event.
This time window is sufficiently wide to allow for full recording of \nai-like scintillation lights ($\tau$=200--300~ns) and also enables us to study large time structures for the photons emitted following  UV exposure. 
%For every 0.15~s of data recording, the DAQ system went offline for approximately 1.35~s in order for the data in the digitizer buffer to be transferred and stored on a DAQ computer. 

All measurements were performed with the crystal module enclosed in the dark box in a darkened room.  When handling of the crystal and PMTs was needed, a red light (640-700~nm) source was used for room illumination, i.e., the source light was not directed to the detector assembly. To expose the crystal to the UV light source (Hg vapor lamp, 365~nm) 
%\jxnote{{explain what UV source}}
, we opened the dark box and removed the top layer of the lead shielding. Parts of the support structure were disassembled, allowing PMT~B to decouple from the \nai\ target (Fig.~\ref{fig_naiSetup}). The optical grease was wiped clean from the PMT~B faceplate and the \nai\ target's quartz window prior to the UV exposure. After UV exposure the optical grease was then reapplied to optically couple the PMT~B and the target. The support structure was reassembled, and the lead shielding was placed back to cover the top of the detector. We maintained a similar PMT-detector orientation before and after UV exposure by indexing PMT~B and the detector, as shown in Fig.~\ref{fig_naiSetup}. After UV exposure, the dark box was closed, and the high voltage power supply was turned on. Data acquisition was then started.
To study the reproducibility of the detector response using the assembly and disassembly procedure, we measured the energy spectra of $^{133}$Ba after repeating this procedure. As shown in Fig.~\ref{fig_ergCali}, variation in the detector response is negligible. We comment that we monitor single PE response during the data-taking period to ensure PMT response stability.

\begin{figure*}[ht!]
    \includegraphics[width=0.97\linewidth]{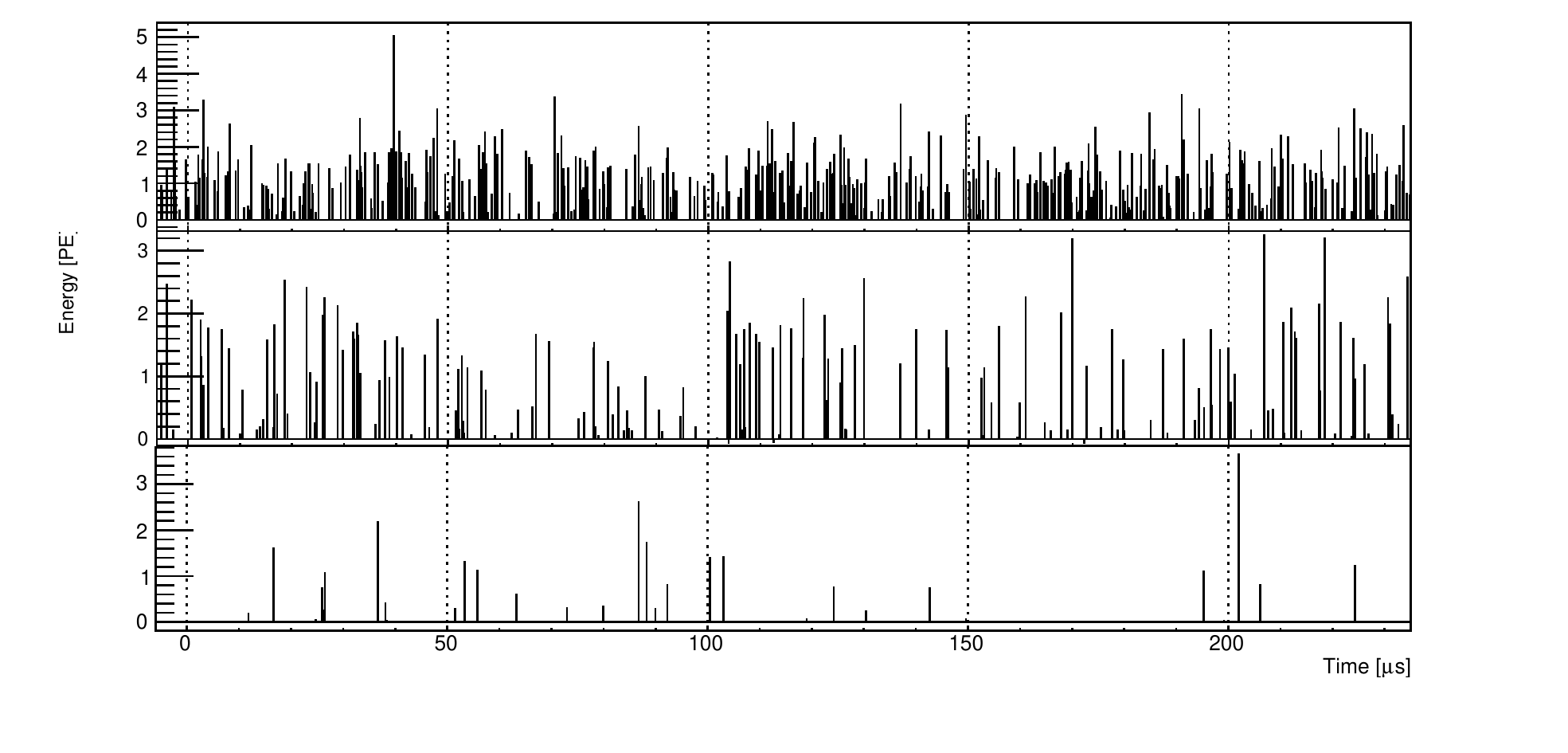} % .png} %pdf}
    \caption{Examples of identified peaks within the 250~\us\ time window 15 minutes, one hour, and two days since the UV exposure.}
    \label{fig_wave250}
\end{figure*}

%%%%%%%%%%%%%%%%%%%%%%%%%%%%%%%%%%%%%%%%%%%%%%%%%%%%%%%%%%%%%%%%%%%%%
\section{\label{sec_discussion}Results and discussion}
%%%%%%%%%%%%%%%%%%%%%%%%%%%%%%%%%%%%%%%%%%%%%%%%%%%%%%%%%%%%%%%%%%%%%

After exposing the NaI(Tl) crystal to a 4-W UV light source (365~nm) from a 20~cm distance for 4 minutes (roughly 1.6$\times 10^{19}$ photons), we observed delayed light emission in the detector that lasted for days.
The intensity of light emission was around 1~PE/$\mu$s shortly after the UV radiation, and it decreased monotonically over time. Similar long-lived delayed light emission was also observed after the \nai\ crystal was irradiated by a $\sim$20~$\mu$Ci $^{60}$Co source from a 3~cm distance for three hours, and after large energy depositions in the crystal ($>$10,000~PE) consistent with passage of muons. 
The observed light emission consists primarily of single photoelectron pulses and does not appear to contain pulse-like structures. 
These observations are consistent with previous reports of light emission in NaI scintillating crystals following particle interactions such as cosmic ray~\cite{delayEmission_cosmicRay_internalRad,delayEmission_cosmicRay_internalRad,delayEmission_cosmicRay_radSource}, radioactive sources~\cite{delayEmission_cosmicRay_radSource}, atmospheric muon~\cite{delayEmission_mu}, and visible lights~\cite{delayEmission_visLight,delayEmission_visLight2} with decay time ranging from $\mu$s to days. 

We rule out the PMTs from being a leading cause of this photoelectron emission because only PMT~A is exposed to the UV light (through the \nai\ crystal module), but both PMT~A and PMT~B observed comparable amounts of photoelectron pulses in this experiment. The quartz windows of the crystal module were exposed to UV light in this experiment, and quartz/fused quartz has been previously reported to emit delayed light lasting from minutes to hours following UV irradiation or X-ray/beta ray exposure in room temperature~\cite{quartz_delayLight1,quartz_delayLight2,quartz_delayLight3,quartz_delayLight4}. However, a report on the afterglow measurement following UV irradiation on alkali halide crystal (KCl) through a quartz window at room temperature suggests that the dominant component in the afterglow is consistent with a known process in the crystal rather than the quartz window~\cite{quartz_carcer}.

%%%%%%%%%%%%%%%%%%%%%%%%%%%%%%%%%%%%%%%%%%%%%%%%%%%%%%%%%%%%%%%%%%%%%%%%%%%%
\subsection{\label{subsec_discussion_tenporal}Waveform characterization}

%\begin{figure}
%    \begin{center}
%        \includegraphics[width=1.0\linewidth]{fig/plot_wave_ba133.pdf}\\
%        (a) \\
%        \includegraphics[width=1.0\linewidth]{fig/plot_fft_ba133.pdf} \\
%        (b) \\
%    \end{center}
%    \caption{The Fourier transform (b) of an example measured NaI(Tl) pulse (a) shows a Lorentzian-like feature, consistent with that expected from exponential decays (signals) in the time domain.}
%    \label{fig_ba133_fft}
%\end{figure}

For the characterization of this delayed photon emission, we use the periodic trigger scheme to avoid trigger biases. 
Peaks were first identified in the recorded raw data containing waveforms from PMT~A and PMT~B, and the charge integrals within each peak were scaled based on the PMT's single photoelectron response. 
The scaled PMT responses were then combined to produce a summed detector event waveform. 
Fig.~\ref{fig_wave250} shows examples of the resulting combined response 15 minutes, one hour, and two days since the UV exposure. 
The width of the bin is set to be 20~ns, which is finer compared to \nai\ pulse decay time.
We comment that 15 minutes is the approximate time required to reassemble the detector support structure after UV radiation is stopped. 

\begin{figure}
    \begin{center}
        \includegraphics[width=1.0\linewidth]{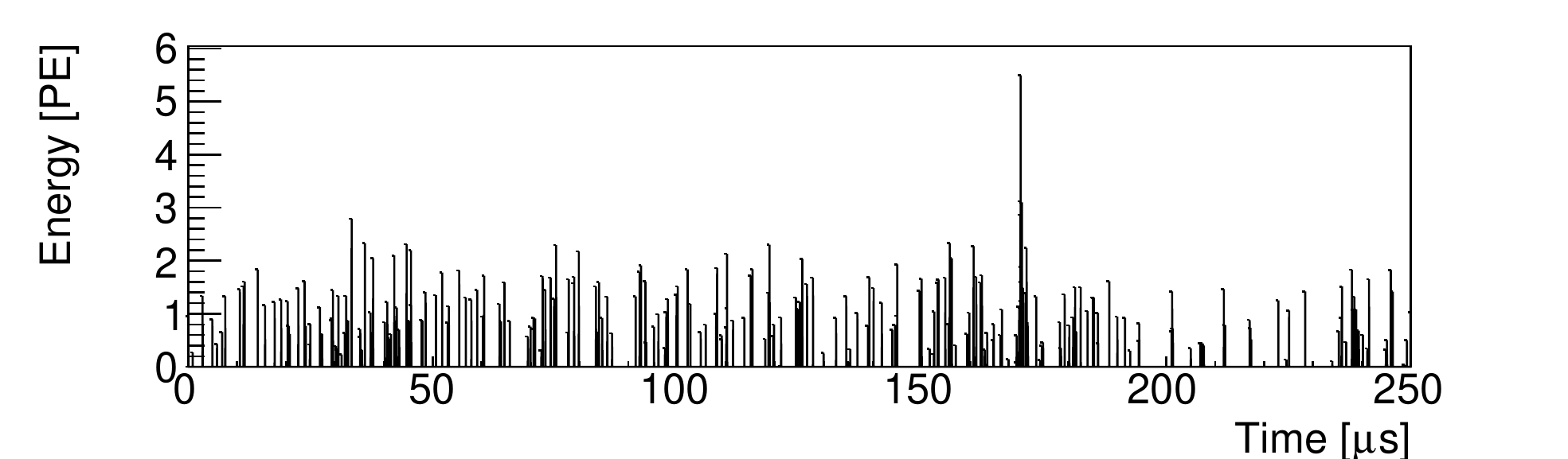} \\ % ave10Wave.pdf} \\ 
        (a) \\
        \includegraphics[width=1.0\linewidth]{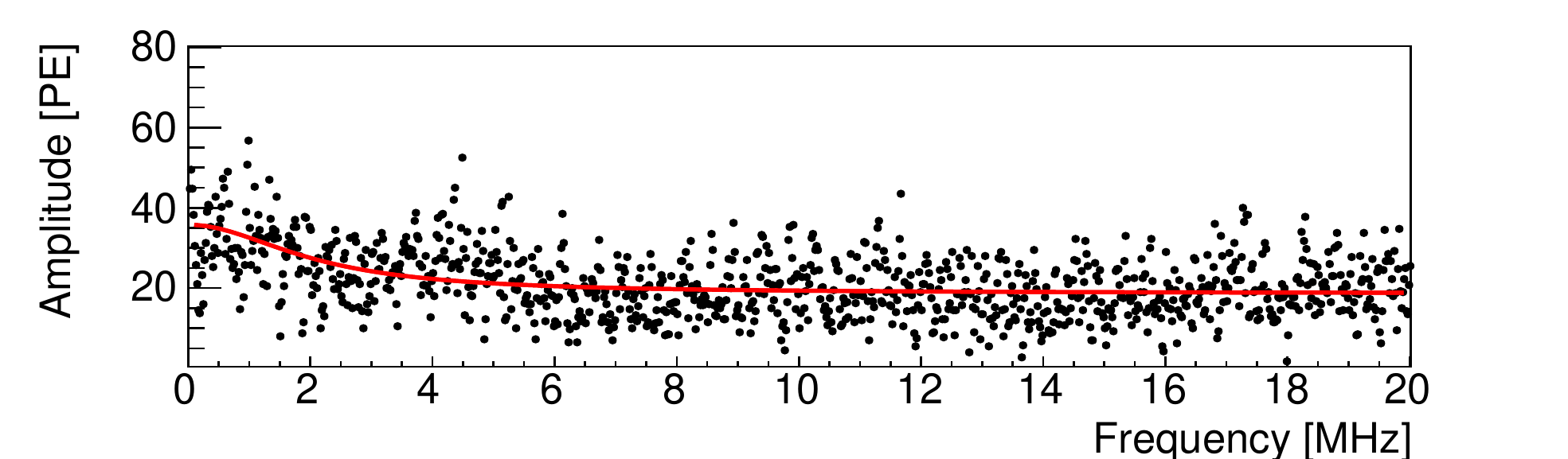} \\ % ave10Wave.pdf} \\
        (b) \\
    \end{center}
    \caption{An example of a simulated waveform with a 2~keV ($\sim$20~PE) NaI(Tl) pulse among relatively intense uncorrelated pulses (1~PE/\us) (a) still shows a noticeable Lorentzian-like feature in the frequency domain (b). %The analysis is applied to ten such a waveform, and the average response in t
    The frequency domain is shown in (b). Smaller simulated pulses produce less noticeable features. The corresponding response in the frequency domain is zoomed from 0 to 10 MHz to show the Lorentzian-like feature. The Lorentzian profile is shown in red.} 
    \label{fig_naiSim_fft}
\end{figure}

\begin{figure}
    \begin{center}
        \includegraphics[width=1.0\linewidth]{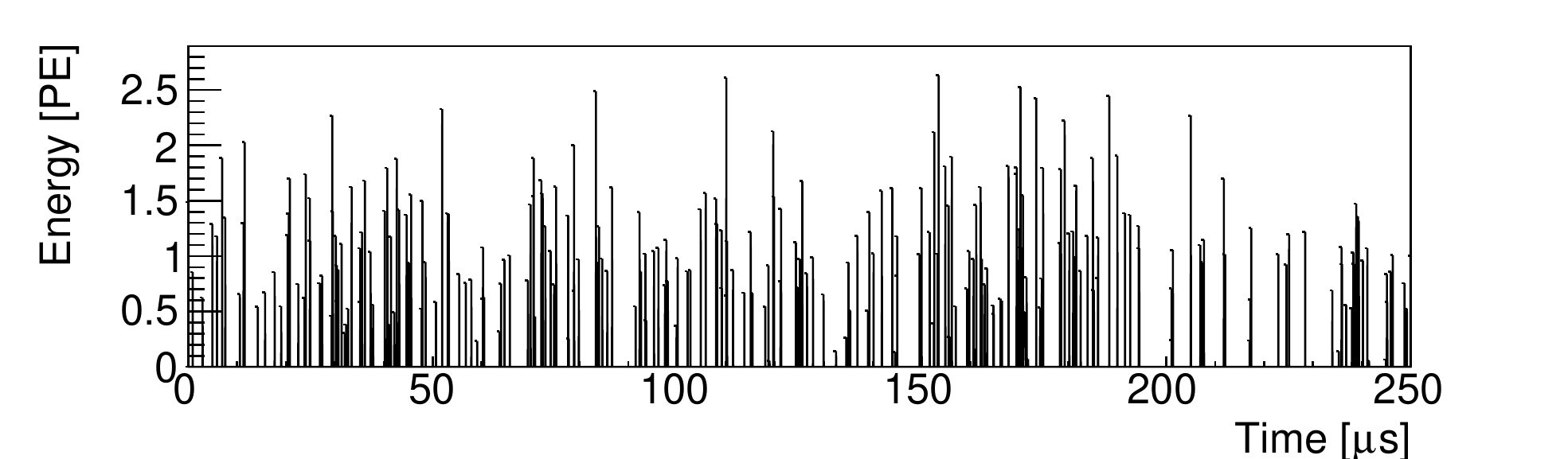} \\ % ave10Wave.pdf}\\
        (a) \\
        \includegraphics[width=1.0\linewidth]{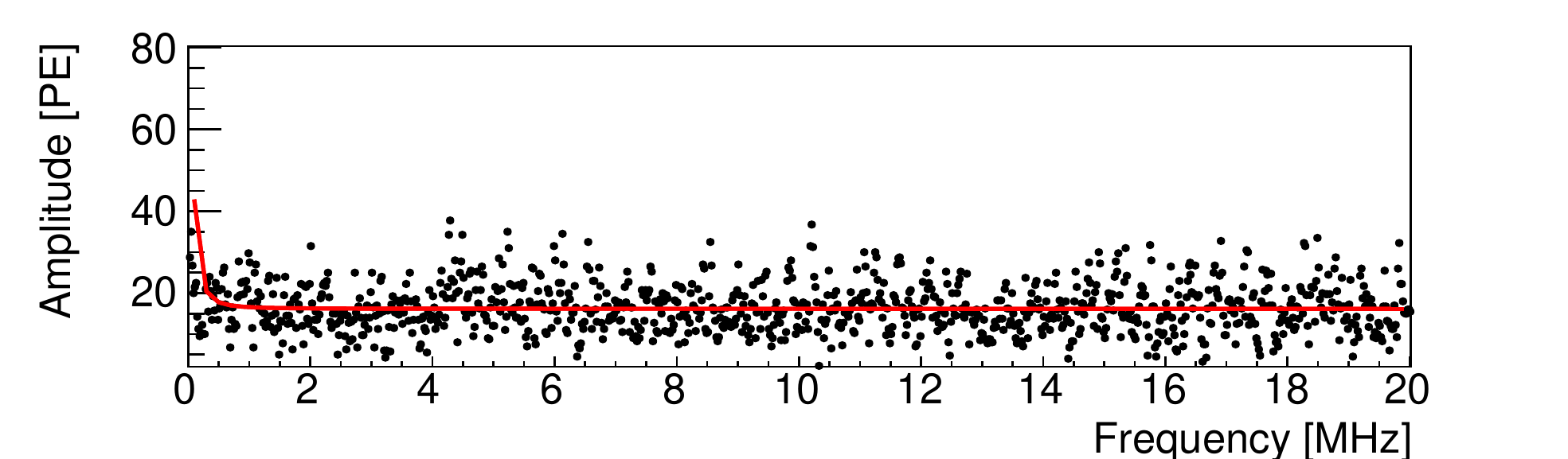} \\ % ave10Wave.pdf} \\
        (b) \\
    \end{center}
    \caption{An example of a simulated waveform populated with uncorrelated pulses (1~PE/\us) (a) shows a uniform response in the frequency domain. The analysis is applied to ten such a waveform, and the average response in the frequency domain is shown in (b). The corresponding response in the frequency domain is zoomed from 0 to 10 MHz to show the absence of the Lorentzian-like feature (red).}
    \label{fig_random_fft}
\end{figure}

%To search for possible features in the continuous flow of photoelectrons,  we apply a Fast Fourier transform (FFT) analysis to the acquired event waveforms. 
We apply Fast Fourier transform (FFT) analysis to the acquired event waveforms to identify their frequency components. 
FFT is a mathematical operation that decomposes a signal in the time domain into an alternate representation in the frequency domain~\cite{fft_korner,fft_morin}.
A random flow of uncorrelated photoelectron pulses would produce a flat frequency spectrum (aka white noise spectrum), while clustered photoelectron structures may demonstrate themselves as nontrivial features at characteristic frequencies.
For the case of \nai\ scintillation pulses, which can be modeled as an exponential decay with a characteristic time constant, the FFT should produce a Lorentzian-like feature in the frequency domain~\cite{fft_morin,fft_korner}. 
%An example frequency spectrum of a measured \nai\ pulse, calculated using the FFT algorithm implemented in the ROOT data analysis toolkit, is shown in Fig.~\ref{fig_ba133_fft}, where a prominent Lorentzian-like feature can be identified. 
%Note that one side of the response in the frequency domain is a mirror reflection of the other side. 
%This is because, for real signals, the coefficients of positive and negative frequencies become complex conjugates.

To test the power of the FFT analysis to identify \nai\ pulse-like features in a random background-dominated waveform, we performed a toy Monte Carlo simulation to mimic the situation when one 2~keV ($\sim$20~PE) \nai\ scintillation pulse is present in a 250~\us\ time window with the highest observed photoelectron background rate ($\sim$1PE/$\mu$s around 15 minutes after UV radiation). 
First, we generated 250 event times randomly distributed within 250~\us\ time window. Next, one event is treated as a NaI(Tl) scintillation pulse containing 20 single-photoelectron (SPE) peaks following an exponential decay with 250~ns time constant, and the rest of 249 events are simply SPE pulses.
For each simulated PE pulse, the height was sampled from the measured SPE Gaussian distribution with a mean of 211~ADC counts and sigma of 128~ADC counts. 
Figure~\ref{fig_naiSim_fft} shows an example waveform generated by the simulation with 20~ns time bins and the 
%averaged 
corresponding FFT response in the frequency domain.
Note that the amplitude in the frequency domain is the magnitude of the transform ($\sqrt{\text{Re}^{2}+\text{Im}^{2}}$).
%over ten simulated waveforms, where a faint Lorentzian-like feature is noticeable.
The red line in Fig.~\ref{fig_naiSim_fft}(a) corresponds to the Lorentzian profile fitted into the frequency distribution. 
In contrast, when the 2~keV pulse is removed from the simulations, the FFT spectrum becomes a flat white noise response, as illustrated in Figure~\ref{fig_random_fft}.
The Lorentzian profile fitted to the frequency distribution (Fig.~\ref{fig_random_fft}(b), red line)  does not show the expected shape.
%The red lines in Figs.~\ref{fig_naiSim_fft} and~\ref{fig_random_fft} are the fits using the Lorentzian profile.

\begin{figure*}[ht!]
    \includegraphics[width=0.94\linewidth]{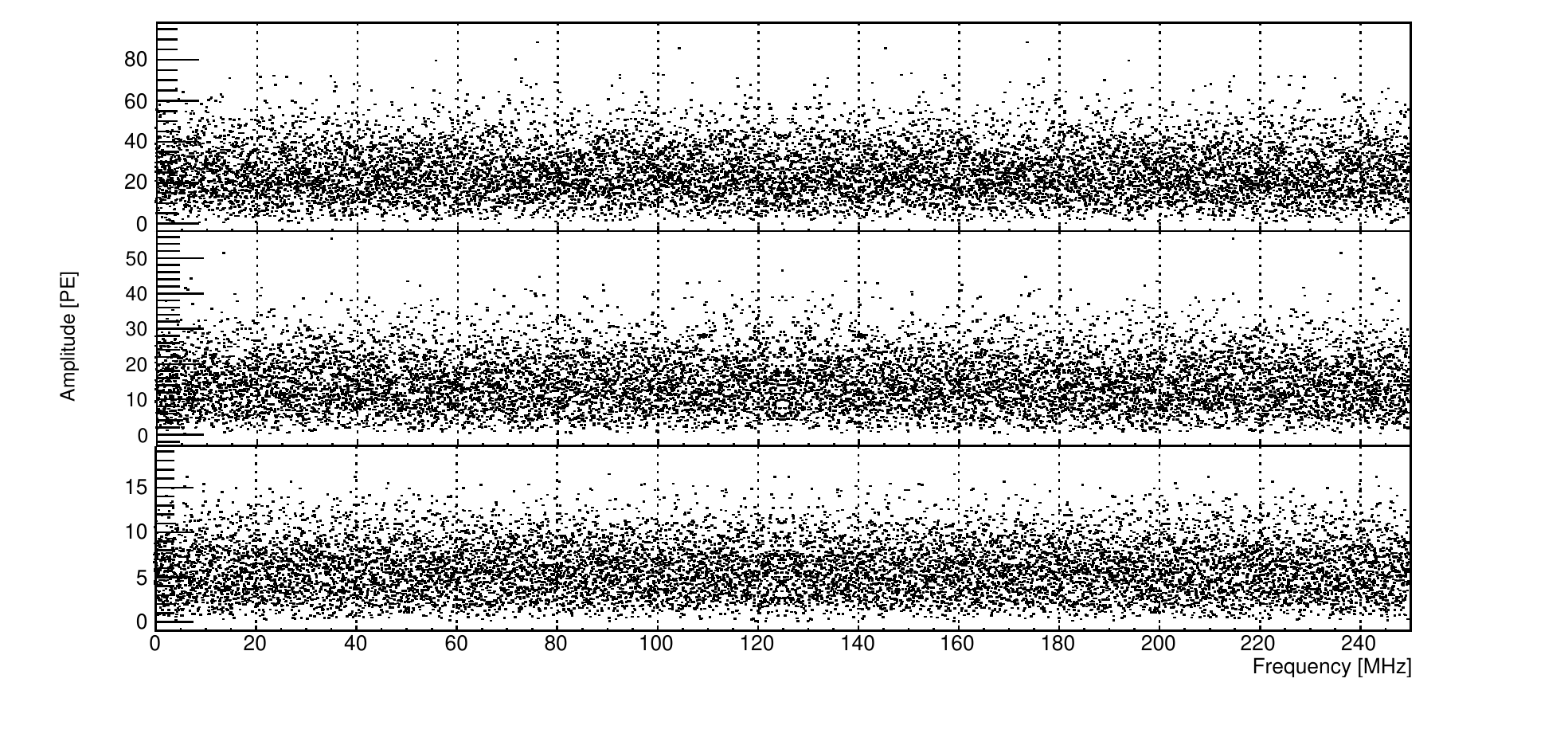}
    \caption{Fast Fourier Transform of NaI(Tl) energy response 15 minutes, one hour, and two days since the UV exposure shows uniform distribution, suggesting that most of the photoelectrons are uncorrelated with each other.}
    \label{fig_fft}
\end{figure*}

The FFT spectra 
%averaged over ten event waveforms 
measured at 15 minutes, one hour, and two days after the UV exposure (example waveforms are given in Figure~\ref{fig_wave250}) are shown in Fig.~\ref{fig_fft}.
The absence of features in the frequency domain, in combination with the toy Monte Carlo simulation study, suggests that the observed photoelectrons following UV exposure are dominated by uncorrelated single-photoelectron pulses. 

We acknowledge that the FFT analysis has a finite sensitivity and can be obscured if the pulse-like events have average energy significantly below 2~keV or if the pulse occurrence is substantially below one per 250~\us\ as we assumed in the toy simulation. 
Therefore, this analysis does not rule out the Saint Gobain report of few-keV pulse-like events (contain time-correlated photons by definition) produced by UV light. 
Request to Saint Gobain for more technical information on their definition of pulse-like events in their measurement and how the rate of these events was calculated was unfruitful. 
A more in-depth analysis of our recorded data may be able to identify occasional few-keV events that are in excess to radiation interactions in the \nai\ crystal and are masked by the high rate of uncorrelated single photoelectron events but is beyond the scope of this study. For this work, we simply report that the substantial and dominant emission of single-photoelectron events observed following UV exposure are uncorrelated.

\subsection{\label{subsec_red}The delayed emission rate and its suppression}

\begin{figure}[ht!]
    \includegraphics[width=1.0\linewidth]{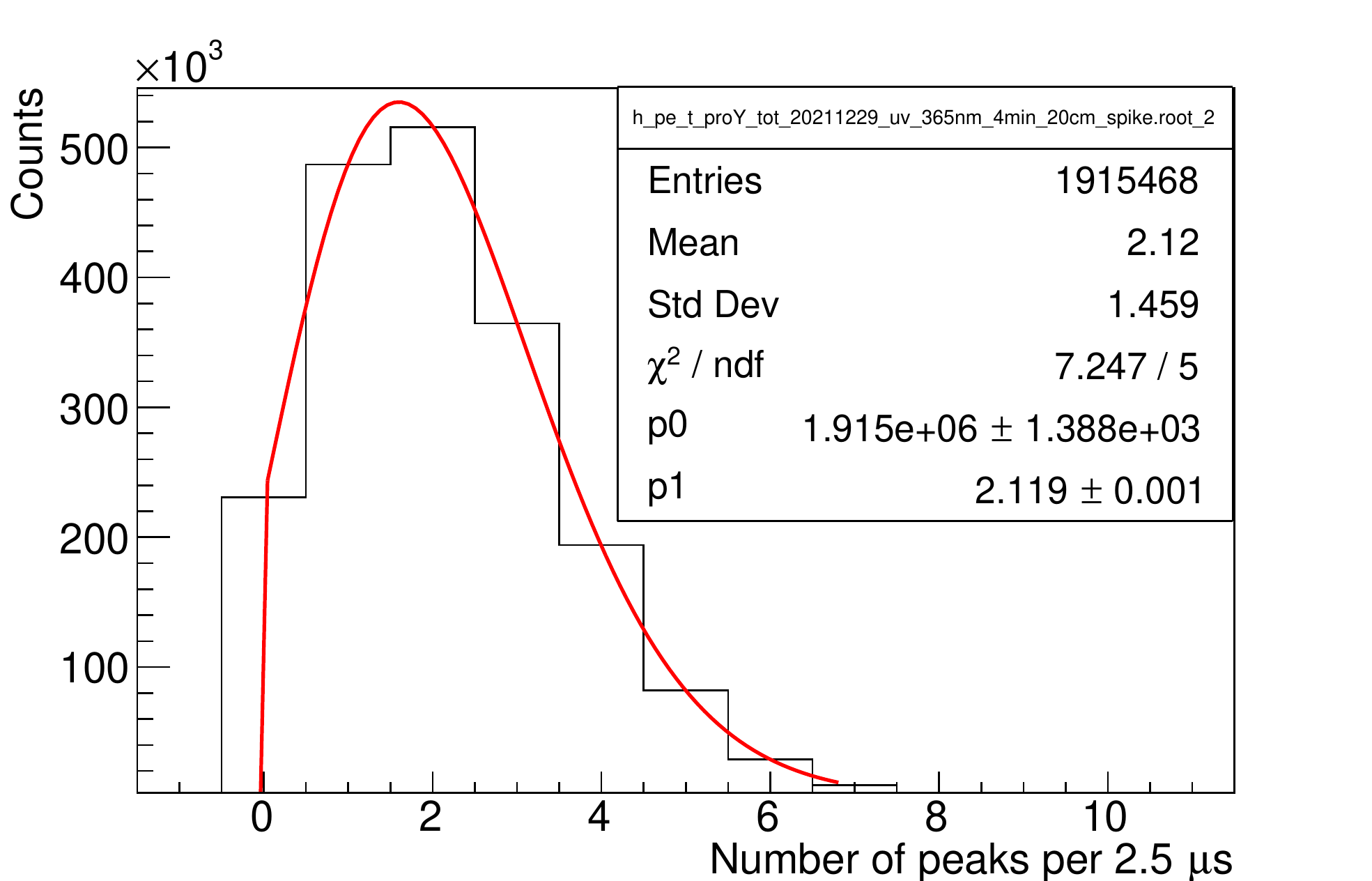}
    \caption{Example of the distribution of the number of observed photoelectron peaks 45 minutes to one hour after the UV exposure.}
    \label{fig_poissonPeak}
\end{figure}

In this section, we characterize the rate of the delayed photon emission from the \nai\ crystal following various forms of radiation.  
For this study, the coincidence trigger, as detailed in Section.~\ref{sec_materialAndMethod}, was used to reduce the data rate. 
To mitigate bias from the coincidence trigger requirement, we exclude the waveform segments both before and around the trigger time in the analysis and only focus on the time window 1.5~$\mu$s to 4.0~$\mu$s after the trigger time. 
Because the waveforms are dominated by the uncorrelated flow of photoelectron pulses, recorded activities in this post-trigger time window approximate that in a triggerless scenario. 
We comment that the number of observed photoelectron peaks in this post-trigger time window approximately follows a Poisson distribution, which further confirms the uncorrelated feature of these photoelectron features. Fig.~\ref{fig_poissonPeak} shows an example of the distribution of the number of observed photoelectron peaks 45 minutes to one hour after the UV exposure. The Poisson profile (Fig.~\ref{fig_poissonPeak}, red line) used to fit the distribution shows respectable $\chi^{2}$/NDF.

\begin{figure}
    \begin{center}
        \includegraphics[width=0.95\linewidth]{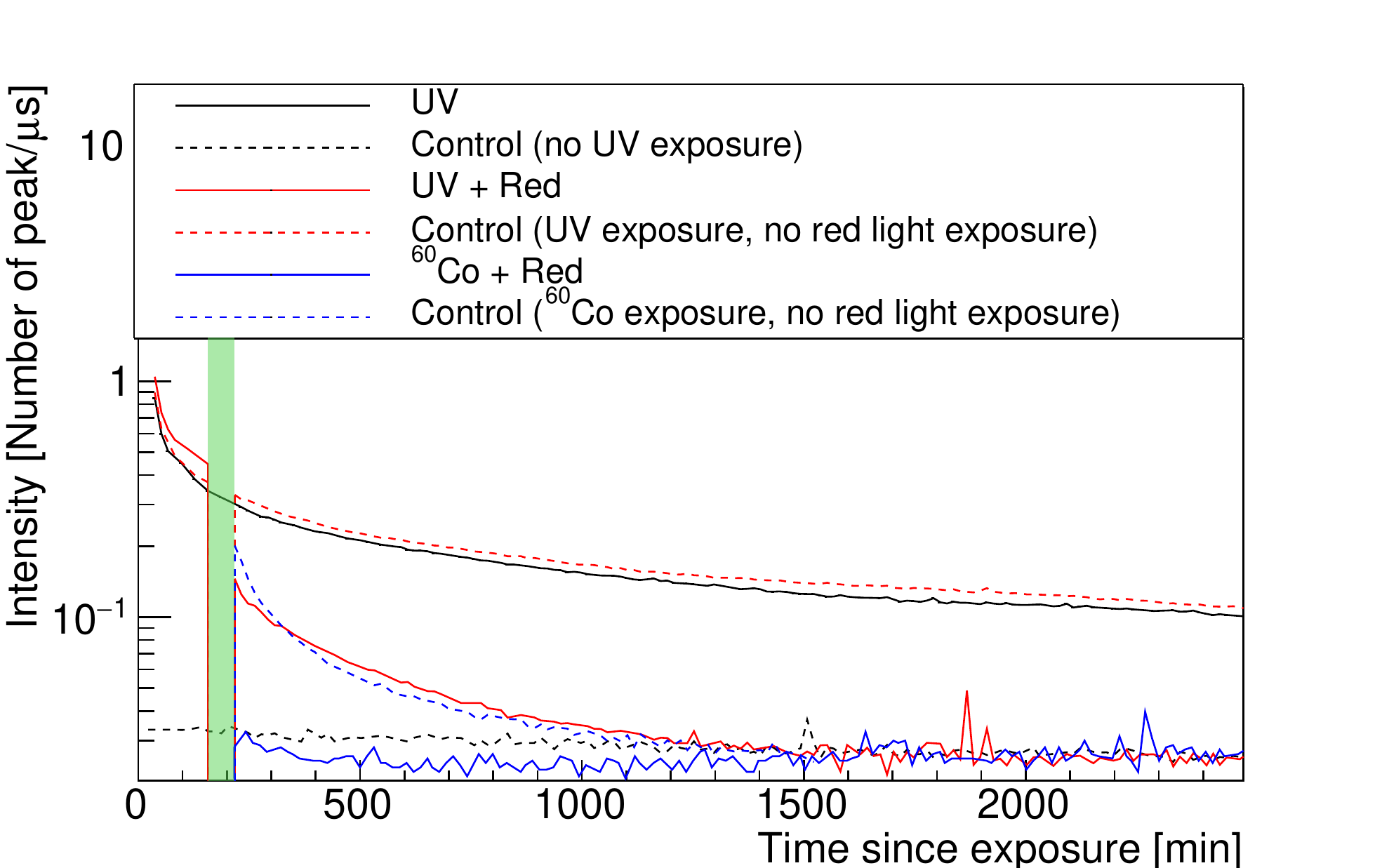} \\% 3.pdf} \\
        (a) \\
        \includegraphics[width=0.95\linewidth]{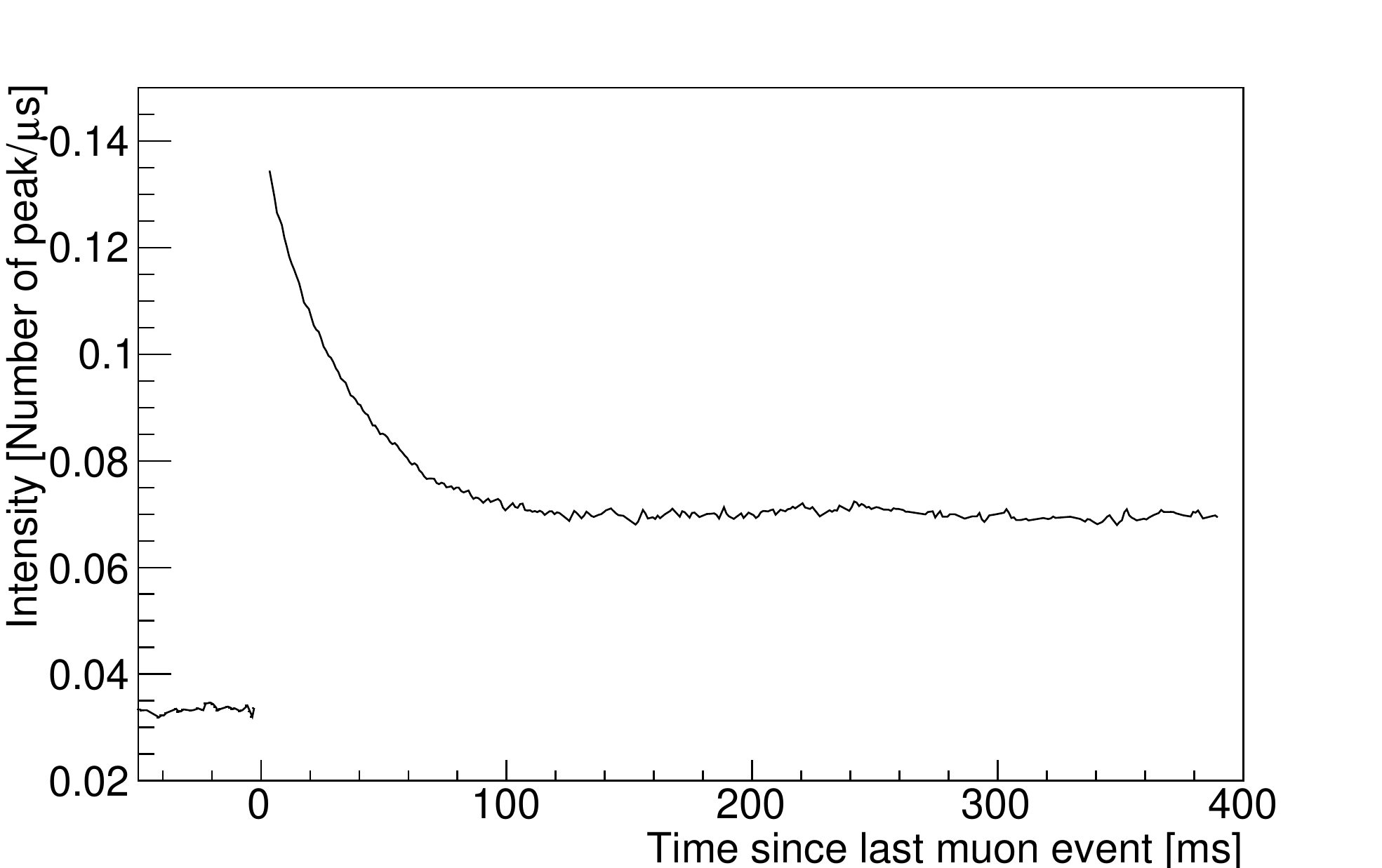} \\ % \\
        (b) \\
    \caption{The delayed light intensity was monitored over time following  UV exposure (a, solid black line) and Co-60 irradiation (a, solid blue line). Three hours after the UV exposure, the crystal was exposed to red light (a, solid red line); and a pronounced decrease in the afterglow intensity was observed. The shaded green box corresponds to the red light irradiation time. Control runs were performed, in which exposure with the UV or red light sources was omitted, but the same procedure was used for source placement as in the runs with exposure. This check confirmed that the experimental procedure had a negligible contribution to the observed trend (a, dashed lines). We also observed delayed light following large energy depositions in the crystal (b). Note that we omit the data point at the zero time, which is the time of the large ionization event (b).}
    \label{fig_intensityTime}
    \end{center}
\end{figure}

Fig.~\ref{fig_intensityTime}(a) shows the mean number of observed photoelectron peaks in the post-trigger window from Poisson fits as a function of time. 
The intensity of light emission in the \nai\ crystal decreases monotonically over time, with an observed photoelectron rate of $\sim$1~PE/$\mu$s shortly after the UV radiation, which decreases to less than 0.1~PE/$\mu$s after a day. 
We comment that the UV experiment was repeated and resulted in a broadly similar trend, demonstrating the reproducibility of the result. 
A control experiment, in which the same detector disassembly and reassembly procedure was exercised but without the UV light source turned on, was also performed.
In this test, no increase in photoelectron rate was observed (Fig.~\ref{fig_intensityTime}, a), confirming that the experimental procedure does not contribute to the delayed light emission.

Following $^{60}$Co radiation, the photon background rate decreases faster than that observed for UV radiation, and the rate returns to the baseline after several hours.
Between subsequent measurements, the \nai\ detector is left in the dark box for at least a week to allow the delayed light emission rate to return to the baseline and not contaminate later measurements.
%In the case of single muons passing the crystal, enhanced light emission can be observed for hundreds of milliseconds (Fig.~\ref{fig_intensityTime}, b).
In the case of single muons passing the crystal, we search for large ionization events ($>$10,000~PE) in the background data (Fig.~\ref{fig_intensityTime}(a), dotted black line) and monitor the intensity of the number of observed photoelectrons before and after the large ionization event. The result is shown in Fig.~\ref{fig_intensityTime}(b). Note that we exclude time zero, which corresponds to the time of the large ionization event, resulting in the observed discontinuity observed in Fig.~\ref{fig_intensityTime}(b).
It has been reported that properties of delayed light emission in alkali halide, such as intensity, spectral composition, and decay time, are affected by the purity of raw material, heat treatment, and imparted radiation doses~\cite{processInorganicScint}. Hence, the intensity, irradiation time, and radiation modes may alter the observed decay time of the luminescence.

While the precise mechanisms of delayed light production by \nai\ following UV radiation are not well understood at present, long-lived metastable activator states and trapping due to UV-induced crystal defects can be partly responsible~\cite{processInorganicScint}.
The former refers to an activator center (Tl) excited to a forbidden state, in which case additional energy is required to promote the activator to another state that is allowed to decay. The additional energy is typically gained from thermal excitation. The resulting delayed light emission hence follow the metastable activator state lifetime and are postulated to range from $\mu$s to days.
The latter refers to excitons that are immobilized by lattice distortion field~\cite{ste_william} and also require additional energy to hop from site to site. 
In both cases, external energy injection into the system can expedite the decay of the metastable and trapped states and thus suppress long-lived light emission. 

As suggested in~\cite{sergey_2021}, we next tested injecting energy into the \nai\ crystal by radiating it with red light after UV exposure to study how it may change the delayed photon emission. 
First, we exposed the \nai\ crystal to UV light (365~nm, 4~W, 4~min, 20~cm distance) as described in Sec.~\ref{subsec_discussion_tenporal} and monitored the light emission for approximately 3 hours. 
Then the PMT high voltage supplies were turned off, the detector was disassembled, and the crystal was bathed in red light (640--700~nm) from a Dewalt DCL043 unit (20~W) for 30 minutes. 
The absorption length of red light in NaI(Tl) is calculated to be at the order of tens of cm~\cite{nai_red_abslength}, so the entire crystal volume (3 in.) was effectively exposed to the red light in this process. 
Afterward, the detector was assembled again, and the crystal scintillation response continued to be monitored by the DAQ. 
Such a procedure means that the measured intensity would drop to zero during the red light irradiation (Fig.~\ref{fig_intensityTime} (a), shaded green region).

Fig.~\ref{fig_intensityTime} (a) shows the delayed photon intensity following UV radiation with and without red light exposure; a large drop in the light intensity is observed after the red light exposure, and the residual light emission rate also decreases at a faster pace. 
A control experiment was carried out, in which the identical procedure was followed only without the red light turned on after the detector was disassembled.
In the controlled experiment, the delayed light emission rate appears to follow a continuous decreasing trend with no abrupt change following the detector disassembly and reassembly procedure. 
Simultaneous exposure to red light and $^{60}$Co irradiation shows a similar decrease in the delayed light emission intensity, as shown in Fig.~\ref{fig_intensityTime}(a) (solid blue line).
We note that the idea of energy injection via longer-wavelength photon irradiation to force the release of trapped states also has been proposed for other detectors. For example, the effect of infrared light irradiation to release trapped electrons by electronegative impurities in a liquid Xenon TPC detector was investigated but the effect was found to be minimal~\cite{purdue_IR}. To the best of our knowledge, our result is the first demonstration of luminescence reduction following UV and gamma radiation via red light exposure in \nai\ detector.

%%%%%%%%%%%%%%%%%%%%%%%%%%%%%%%%%%%%%%%%%%%%%%%%%%%%%%%%%%%%%%%%%%%%%%%%%%%%
\subsection{\label{subsec_otherExp}Relevance to NaI(Tl)-based dark matter experiments}

The DM-Ice17 collaboration extensively studied the muon-induced delayed light emission and whether it can mimic the \dl\ modulation~\cite{dm-ice_muon,hubbard_dissertation}. They concluded that this process is unlikely to be the sole source of the \dl\ modulation for two reasons. First, the event rate is too low compared to the \dl\ observation (even neglecting noise suppression cuts in the analysis), and second, the phase of the muon flux annual pattern is off by approximately 45~days compared to the \dl\ modulation. 
However, it is possible that other forms of energy storage in the \dl\ crystals from both internal and external radiation (such as alpha decays from U/Th chains), in combination with different sources of annual variation~\cite{alterPhase} include temperature, pressure, vibration, electric field~\cite{ki_elecField} in addition to muon flux, may still play a role in the \dl\ experiment. 
We also comment that delayed light emission may appear as a single-hit event or a multiple-hit event depending on the type of particles inducing the delayed light production. For example, muon passage across an array of \nai\ crystals may result in multiple-hit events. In contrast, short-ranged particles like alpha likely produce single-hit events.

\begin{figure}[ht!]
	\begin{center}
    \includegraphics[width=0.5\textwidth]{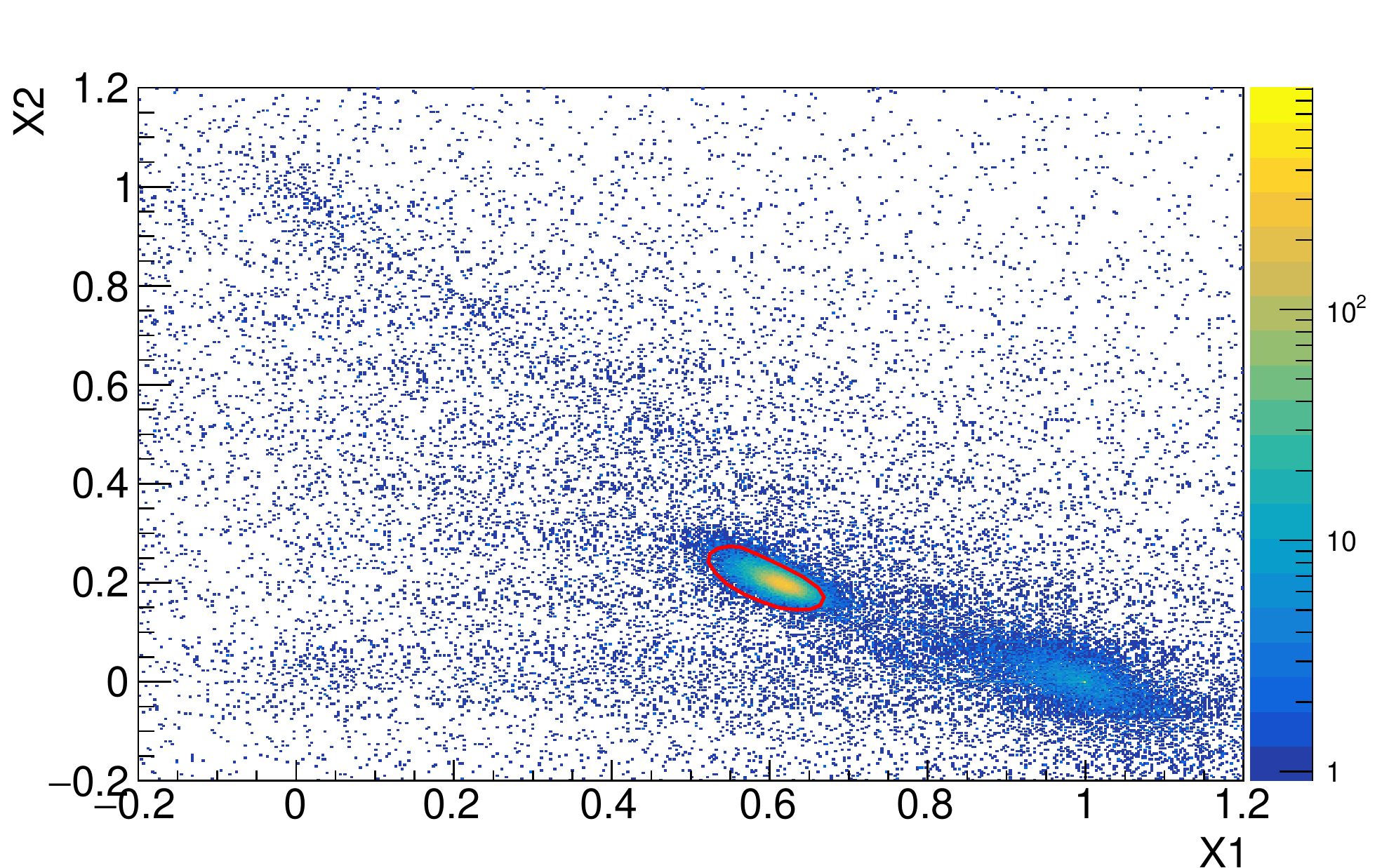}\\
    (a) \\
    \includegraphics[width=0.5\textwidth]{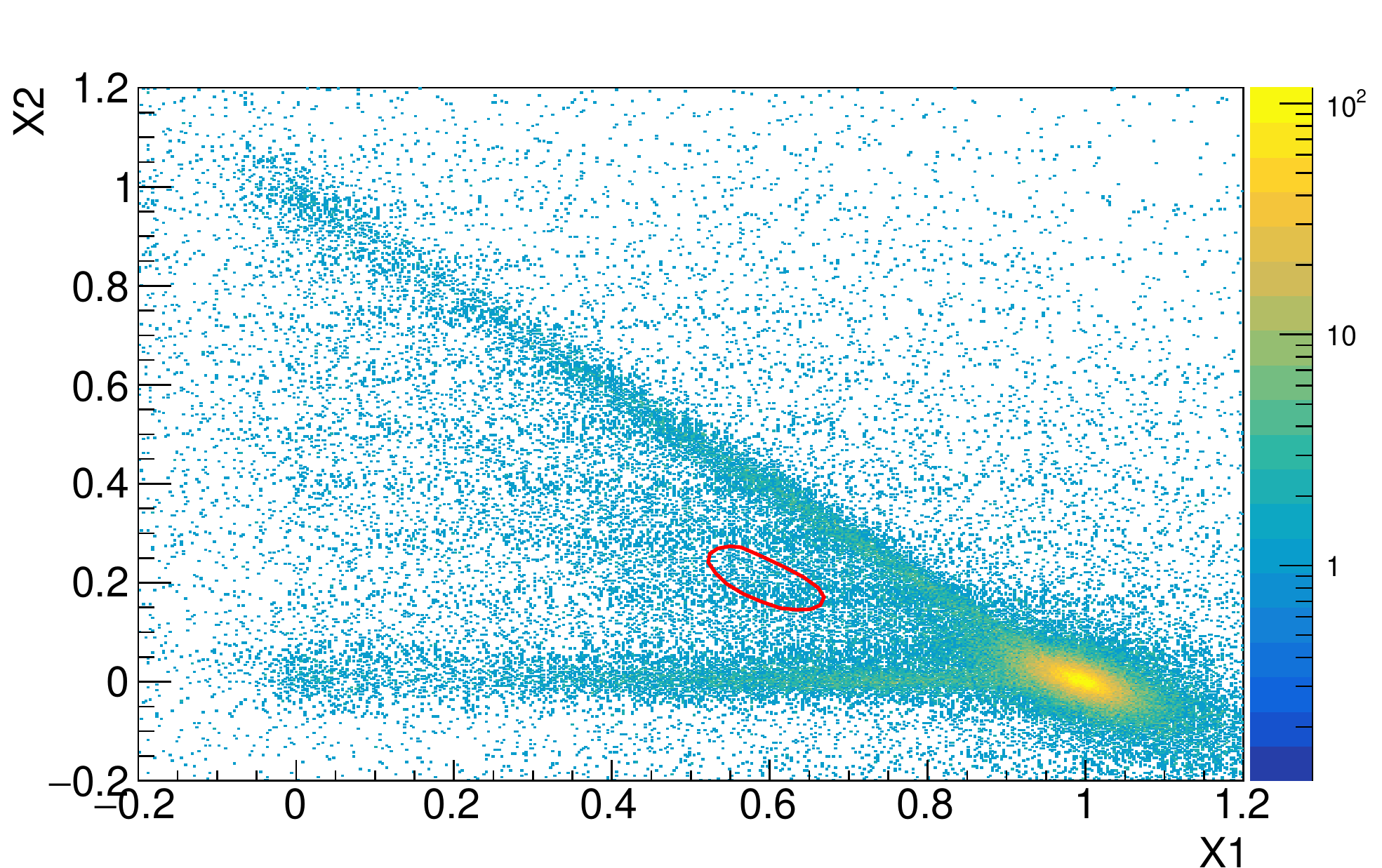}\\
    (b)
	\end{center}
    \caption{$^{133}$Ba data reveals the position of traditional NaI(Tl) pulse (red circle) in the $X_1$-$X_2$ parameter space (a). UV-induced delayed light emission has the potential to leak into the $X_1$-$X_2$ region of genuine NaI(Tl) pulse (b).  }
    \label{fig_psd}
\end{figure}

%Although we did not observe pulse-like background events introduced by UV light as reported by Saint Gobain, we cannot definitively refute this claim due to the excessive photon emission background. 
Here we consider the possibility of radiation-induced delayed light emission leaking through some \dl-like pulse shape-based event selection cuts. 
A train of uncorrelated single photoelectrons can pass the trigger coincidence requirement commonly employed by dark matter experiments, and these events may further pass analysis cuts designed to select \nai\  scintillation pulses. 
\dl\ defines pulse shape parameters, $X_1$ and $X_2$, or the fraction of delayed and prompt pulse integrals, respectively, to discriminate \nai\ scintillation events from noise-like events~\cite{dama-libra-apparatus}.
\begin{equation} \label{eq_x1x2}
    X_1 = \frac{A[100,600]~ns}{A[0,600]~ns}, 
    X_2 = \frac{A[0,50]~ns}{A[0,600]~ns}
\end{equation}
Because \nai\ scintillation pulses have a relatively long decay time, while PMT noise events that pass the coincidence requirement are more likely to have short durations, a genuine \nai\ scintillation event is expected to have a large $X_1$ and a relatively low $X_2$ value. 

In this exercise, we choose an event window of 2048~ns after the trigger time, which is the same as the \dl's time recording window. 
To simulate the effect of additional pulse-selection cuts, we also require a candidate event to have an integrated charge less than 10\% of the total event energy in the pre-trigger region [-1,-0.5]~$\mu$s and at a later time window [1,1.5]~$\mu$s.
Additionally, we only accept events within 1--6~keV energy regions.
We took a $^{133}$Ba data to show the expected $X_1$-$X_2$ response to genuine \nai\ pulses (Fig.~\ref{fig_psd}(a)).
The red circle corresponds to the region where genuine \nai\ pulses are located. 
The heavy population in the ($X_1$,$X_2$)$=$(1,0), the horizontal line at $X_2=$0, and the diagonal line extending from ($X_1$,$X_2$)$=$(1,0) to ($X_1$,$X_2$)$=$(0,1) are expected from dark counts. 
Since the UV-induced delayed light emission has a dominant uncorrelated component, the resulting $X_1$-$X_2$ response mimics that of dark counts, as shown in Fig.~\ref{fig_psd}(b).
Although the $X_1$-$X_2$ responses are significantly different for genuine \nai\ pulses produced by gamma-ray interactions and for delayed light emission, the UV-induced background is capable of contaminating the region where genuine \nai\ pulses are located, as shown in  Fig.~\ref{fig_psd}(b). 
Note that this is simply an illustrative analysis, and the contamination level may vary from one experiment to another due to the variations in the crystal properties, data acquisition methods, and analysis cuts. 
If a \nai-based dark matter experiment observes leakages of backgrounds in this manner, they would need to characterize the leakage fraction for each crystal and consider suppressing this parasitic background by injecting small amounts of energy into the crystal to force energy relaxation, as demonstrated in Sec.~\ref{subsec_red}.

%%%%%%%%%%%%%%%%%%%%%%%%%%%%%%%%%%%%%%%%%%%%%%%%%%%%%%%%%%%%%%%%%%%%%
\section{\label{sec_conclusion}Conclusions}
%%%%%%%%%%%%%%%%%%%%%%%%%%%%%%%%%%%%%%%%%%%%%%%%%%%%%%%%%%%%%%%%%%%%%

We exposed a 3'' \nai\ crystal to a 365~nm UV light source and observed strong light emission that lasted for multiple days, which may be explained by energy accumulation and delayed release in the detector. 
Gamma-ray irradiation and muon events are also observed to produce similar phenomena.
We found that uncorrelated photons dominate such light emission, but this background could still pass a coincidence trigger and be mistakenly categorized as keV-scale \nai\ scintillation events by \dl-style pulse shape selection cuts.
These observations may prompt reconsideration of seasonal-modulated external sources as a possible explanation for the \dl\ signal. For example, muon flux, changes in temperature, vibration due to the nearby traffic, pressure variation, and electric/magnetic field~\cite{ki_elecField} exhibit yearly patterns, and a combination of external energy inputs may yield apparent modulation similar to that observed in \dl. 

We found that red light exposure suppresses the intensity of delayed light emission in \nai\ by stimulating the dissipation of stored energy in the system. However, the use of red light in an actual experiment may be difficult due to the finite sensitivities of common photocathodes in the red light region. Future work should investigate the potential of longer-wavelength (IR or mid-IR) sources to suppress the production of delayed light. Other stimuli, such as small but well-controlled temperature or pressure changes, or vibrations, may also impact the accumulation and dissipation of stored energy in NaI(Tl) detectors or other systems. 

%This work could not definitively confirm or refute the Saint-Gobain claim of keV-scale events produced in NaI(Tl) crystals by UV exposure due to the high light emission background. 
%While this work could not definitively confirm or refute the Saint-Gobain claim of keV-scale events produced in NaI(Tl) crystals by UV exposure due to the high light emission background, we report two important observations:
%\begin{itemize}
%\item UV-induced delayed light emission has a dominant uncorrelated component, a piece of information missing from the Saint Gobain report. 
%\item Red light irradiation can suppress potential parasitic background due to the uncorrelated emission of delayed light.
%\end{itemize}
%Further studies with improved sensitivities may determine whether few-keV equivalent pulses of very low rates can be induced via UV or other irradiation. 

%%%%%%%%%%%%%%%%%%%%%%%%%%%%%%%%%%%%%%%%%%%%%%%%%%%%%%%%%%%%%%%%%%%%%
\section{Acknowledgements}
%%%%%%%%%%%%%%%%%%%%%%%%%%%%%%%%%%%%%%%%%%%%%%%%%%%%%%%%%%%%%%%%%%%%%

We thank D. Nygren and F. Calaprice for their helpful discussion on the potential contribution of accumulated energy discharge to the observed DM signal in the \dl\ experiment.
We also thank Antonia Hubbard for insightful discussions on the muon-induced delayed light emission in NaI(Tl) crystal. 
This project is supported by the U.S. Department of Energy (DOE) Office of Science/High Energy Physics under Work Proposal Number SCW1508 awarded to Lawrence Livermore National Laboratory (LLNL). 
LLNL is operated by Lawrence Livermore National Security, LLC, for the DOE, National Nuclear Security Administration (NNSA) under Contract DE-AC52-07NA27344 [LLNL release number LLNL-JRNL-837382].

%%%%%%%%%%%%%%%%%%%%%%%%%%%%%%%%%%%%%%%%%%%%%%%%%%%%%%%%%%%%%%%%%%%%%
%\bibliography{biblio}
%%%%%%%%%%%%%%%%%%%%%%%%%%%%%%%%%%%%%%%%%%%%%%%%%%%%%%%%%%%%%%%%%%%%%


\end{document}